\documentclass[aps,prx,twocolumn,superscriptaddress]{revtex4}
\usepackage{graphicx}

\newcommand{\be}{\begin{equation}}
\newcommand{\ee}{\end{equation}}
\newcommand{\ba}{\begin{eqnarray}}
\newcommand{\ea}{\end{eqnarray}}
\newcommand{\baa}{\begin{eqnarray*}}
\newcommand{\eaa}{\end{eqnarray*}}

\def\be{\begin{equation}}
\def\ee{\end{equation}}
\def\bea{\begin{eqnarray}}
\def\eea{\end{eqnarray}}

\def\C60{A$_x$C$_{60}$}

\def\HgCu3{HgCa$_2$Cu$_3$O$_{8+y}$}
\def\HgCu4{HgBa$_2$Ca$_3$Cu$_4$O$_{10+y}$}
\def\TlCu{Tl$_2$Ba$_2$CuO$_{6+\delta}$}
\def\TlCu3{Tl$_2$Ba$_2$Ca$_2$Cu$_3$O$_{10+y}$}
\def\TlCu4{Tl$_2$Ba$_2$Ca$_3$Cu$_4$O$_{12+y}$}

\def\BiCu3{Bi$_2$Sr$_2$Ca$_{2}$Cu$_3$O$_y$}

\def\8LSCO{La$_{1.88}$Sr$_{.12}$CuO$_4$}
\def\110LNSCO{La$_{1.5}$Nd$_{0.4}$Sr$_{0.1}$CuO$_{4}$}
\def\stage4LCO{La$_{2}$CuO$_{4+\delta}$}
\def\Y248{YBa$_2$Cu$_4$O$_8$}

\def\NbSe2{NbSe$_2$}
\def\TaSe2{TaSe$_2$}
\def\TiSe2{TiSe$_2$}

\begin{document}
\title{   Iron-Based Superconductors As Odd Parity Superconductors }
\author{Jiangping Hu}
\affiliation{Beijing National
Laboratory for Condensed Matter Physics, Institute of Physics,
Chinese Academy of Sciences, Beijing 100080,
China}
\affiliation{Department of Physics, Purdue University, West
Lafayette, Indiana 47907, USA}
 
\begin{abstract}Parity is a fundamental quantum number to classify a state of matter.  Materials rarely possess  ground states with odd parity. We show that the superconducting state in iron-based superconductors is classified as an odd parity s-wave spin-singlet pairing state in a single trilayer FeAs/Se, the building block of the materials. In a low energy effective model constructed  on the  Fe square bipartite lattice,  the superconducting order parameter in this state is a combination of a s-wave normal pairing between two sublattices and  a s-wave $\eta$-pairing within the sublattices. Parity conservation was violated in proposed superconducting states in the past.  The  state has a fingerprint with a real space sign inversion between the top and bottom As/Se layers. The results demonstrate  iron-based superconductors being a new quantum state of matter and suggest that  the measurement of the odd parity can establish fundamental principles related to high temperature superconducting mechanism. \end{abstract}
\maketitle
\section{Introduction}

 Symmetry plays the central role  in the search for beauty in physics.  It controls the structure of matter and  allows us  to simplify a complicated problem. Gauge principle is a fundamental principle in physics. Models formulated in different gauge settings are equivalent.  Symmetry and gauge principles together make the foundations of modern physics to solve  complicated problems.

 The recently discovered high temperature superconductors (high-$T_c$), iron-based superconductors\cite{Hosono,ChenXH,wangnl2008}, are layered materials with complicated electronic structures.  The complexity causes a major difficulty in understanding pairing symmetry which arguably is the most important property  and  clue to determine pairing mechanism\cite{Johnstonreview,review-hirschfeld2011}.

In  a strongly correlated electron system, major physics is   determined locally in real space.
 Important properties, such as pairing symmetry in a superconducting state, are expected to be robust against small variation of Fermi surfaces in reciprocal space. Although superconducting mechanism related to high temperature superconductors (high $T_c$)  is still unsettled, the robust d-wave pairing symmetry in cuprates\cite{Tsuei2000} 
can be understood under this principle. 

Is this principle still held for iron-based superconductors? Namely, do all iron-based superconductors possess one universal pairing state?  Unlike cuprates, the answer to this question is highly controversial because different theoretical approaches have provided different answers and no universal state has been identified\cite{review-hirschfeld2011}.  Nevertheless,  as local electronic structures in all families of iron-based superconductors  are  almost identical and phase diagrams  are smooth against doping\cite{Johnstonreview,review-hirschfeld2011}, it is hard to argue that the materials can approach many different superconducting ground states.

In conventional wisdom, there are several obvious requirements regarding  electron pairing in superconducting states. First, pairing symmetry is known to be classified according to lattice symmetry.    Second, in a  uniform superconducting state, the total momentum  for  the Cooper pairs (modulo a reciprocal lattice vector) must be vanished. Finally, for a central symmetric lattice  with a space inversion center, the parity of  superconducting order parameters  normally is even for a spin-singlet pairing and odd for a spin-triplet pairing\cite{Anderson-pair}. These requirements are  easily fulfilled in a system with a simple electronic structure, such as cuprates. However, for iron-based superconductors, they are highly non-trivial.  

 The unit cell in iron-based superconductors is intrinsically a 2-Fe unit cell while  for simplicity, most theoretical models are effectively constructed based on an 1-Fe  unit cell\cite{Kuroki2011,Graser2010,Eschrig2009,leewen2008,daghofer2010,raghu2008}.  Obviously, these effective models have different lattice symmetry from  the models defined on the original  lattice. The difference may cause serious problems.  For example,  in the effective models, the pairings   have been limited to two electrons with opposite momentum $(\vec k, -\vec k)$, which we  call it  normal pairing in this paper, where $\vec k$ is momentum defined with respect to the  1-Fe unit cell in an iron square lattice.  The momentum vector $Q=(\pi,\pi)$   is   a reciprocal lattice vector in the original lattice with a 2-Fe unit cell.  Thus,  the pairings  ( $\vec k, -\vec k+Q$), in principle, are also allowed according to the above second requirements. We will refer this pairing channel as an extended $\eta$-pairing\cite{Yangcn-eta,yang2} and simply call it  $\eta$-pairing in this paper. The possible existence of $\eta$-pairing was discussed in simplified models\cite{Hu2012s4,Gao2010,Khodas2012}.  As order parameters are classified differently under different symmetry groups, we need to understand these orders under  the original lattice symmetry. Otherwise, conseration laws could be violated.
  
In this paper, we  show that the superconducting state in iron-based superconductors is a new state of matter which is classified as an odd parity s-wave spin-singlet pairing state in a single trilayer FeAs/Se, the building block of the materials.   The superconducting states that were proposed  in the past based on the effective d-orbital models of an iron square lattice  are not  parity eigenstates.  Parity conservation was violated.  We show that  the  superconducting state which conserves parity includes both  normal pairing between two sublattices of the iron square lattice and $\eta$-pairing within each sublattices. The  states have a fingerprint with a real space sign inversion between the top and bottom As/Se layers. Our derivation is based on general  symmetry and gauge requirements on the effective models for iron-based superconductors.

In the following,  we first provide a complete symmetry analysis  for  pairing symmetries in iron-based superconductors.  While the pairing symmetries can be classified according to $D_{2d}$  point group at iron sites or $C_{4v}$ point group  at  the center of an iron square,  there are two types of pairing symmetries   for a spin-singlet  pairing state  because of the intrinsic 2-Fe unit cell. They  are  distinguished from each other by opposite parity numbers.  Second, we discuss the hidden symmetry properties of the effective models under the original lattice symmetry. We show that the effective hopping terms between two sublattices and within each sublattice have different symmetry characters. Third, we discuss a general gauge principle related to the definition of pairing symmetries and conclude  that  parity conservation was violated in the past. We  provide the meanfield Hamiltonian in the new superconducting state and show that it can provide a unified description of all families of iron-based superconductors including both iron-pnictides\cite{Johnstonreview} and iron-chalcognides\cite{Guojg2010,Hesl2012,Liudefa2012,Tansy2013}.  Fourth, we discuss the smoking-gun experiments that can reveal the parity of the superconducting state. Finally, we discuss  the fundamental impact on high $T_c$ superconducting mechanism if it is confirmed.

Before we start the main content, we first make clear about a gauge setting for  the effective models that were constructed based on all five  iron $d$-orbitals\cite{Kuroki2011,Graser2010,Eschrig2009}.
   In those effective models  with an 1-Fe unit cell, a new gauge setting  is taken\cite{Kuroki2011,Graser2010,Eschrig2009}, which effectively changes the momentum $\vec k$ to $\vec k+Q$ for $d_{xz}$ and $d_{yz}$ orbitals.   In the following, without further clarification, the momentum $\vec k $  used in the definition of our normal pairing $(\vec k,-\vec k)$ and $\eta$-pairing $(\vec k,-\vec k+Q)$ is the same momentum used in those papers rather than the momentum in a natural gauge setting.  

\section{Symmetry of a Single Fe-As(Se) Trilayer}
  
   Iron-based superconductors are layered materials.  The essential electronic physics is controlled by a single trilayer Fe-As(Se) structure, the building block of the superconductors. Although the coupling along c-axis between the building blocks has many interesting effects,  the superconducting mechanism and the fundamental properties of the superconducting states, such as pairing symmetries,  are expected to be solely determined within  the single building block.  The observation of superconductivity in a single  FeSe layer grown by MBE  has further justified this two-dimensional nature\cite{Hesl2012,Liudefa2012,Tansy2013}.    Therefore,   we first focus on the study of  a  single   Fe-As(Se) trilayer structure.

We start our analysis  by  understanding the lattice symmetry.  As shown in Fig. 1, 
 the structure has an inversion center, the origin,  located  at  the middle of each Fe-Fe link. The unit cell with the origin at the center is marked by the shadowed area, which includes two irons and two As/Se atoms.  We denote $T$ as the translation group with respect to the unit cell.  The symmetry group,  thus, is described by a non-symmorphic space group $G=P4/nmm$\cite{Fischer2011}.  The quotient group $G/T$ is specified by 16 symmetry operations that include a space inversion $\hat I$. It is easy to check that these operations can be specified equivalently as  $C_{4v}\oplus \hat I C_{4v}$ or $D_{2d}\oplus \hat I D_{2d}$, where $C_{4v}$ is the point group with respect to the point at the middle of an iron square and $D_{2d}$ is the point group defined at an  iron site. It is important to note that both $C_{4v}$ and $D_{2d}$ are not defined with respect to the inversion center. Therefore, some symmetry operations in $C_{4v}$ or $D_{2d}$ are non-symmorphic. For example,  the $\hat C_4$ rotation operation in $C_{4v}$ is  equivalent to   $(\hat C'_{4}, \hat t_2)$, which represents $\hat C'_{4}$, an rotation $\frac{\pi}{2}$ along z-axis at the inversion center, followed by a translation operation $\hat t_2$ that translates $(\frac{1}{2},\frac{1}{2},0)$ in the coordinate of an iron square lattice, a half of unit lattice cell along $X$ direction labeled in Fig.1.
 
 In summary, the full symmetry group  can be written as
 \begin{eqnarray}
 G/T= Z_2\otimes D_{2d} =Z_2\otimes C_{4v}
 \end{eqnarray}
 where $Z_2=(\hat E, \hat I)$.  The group  is  a direct product of two subgroups which are defined with respect to different operation centers.  $\hat I$ commutates  with all symmetry operations in $D_{2d}$  or $C_{4v}$ in a sense that operations are considered to be identical if they only differ by a lattice transition with respect to the 2-Fe unit cell.

\begin{figure}
\begin{center}
\includegraphics[width=1\linewidth]{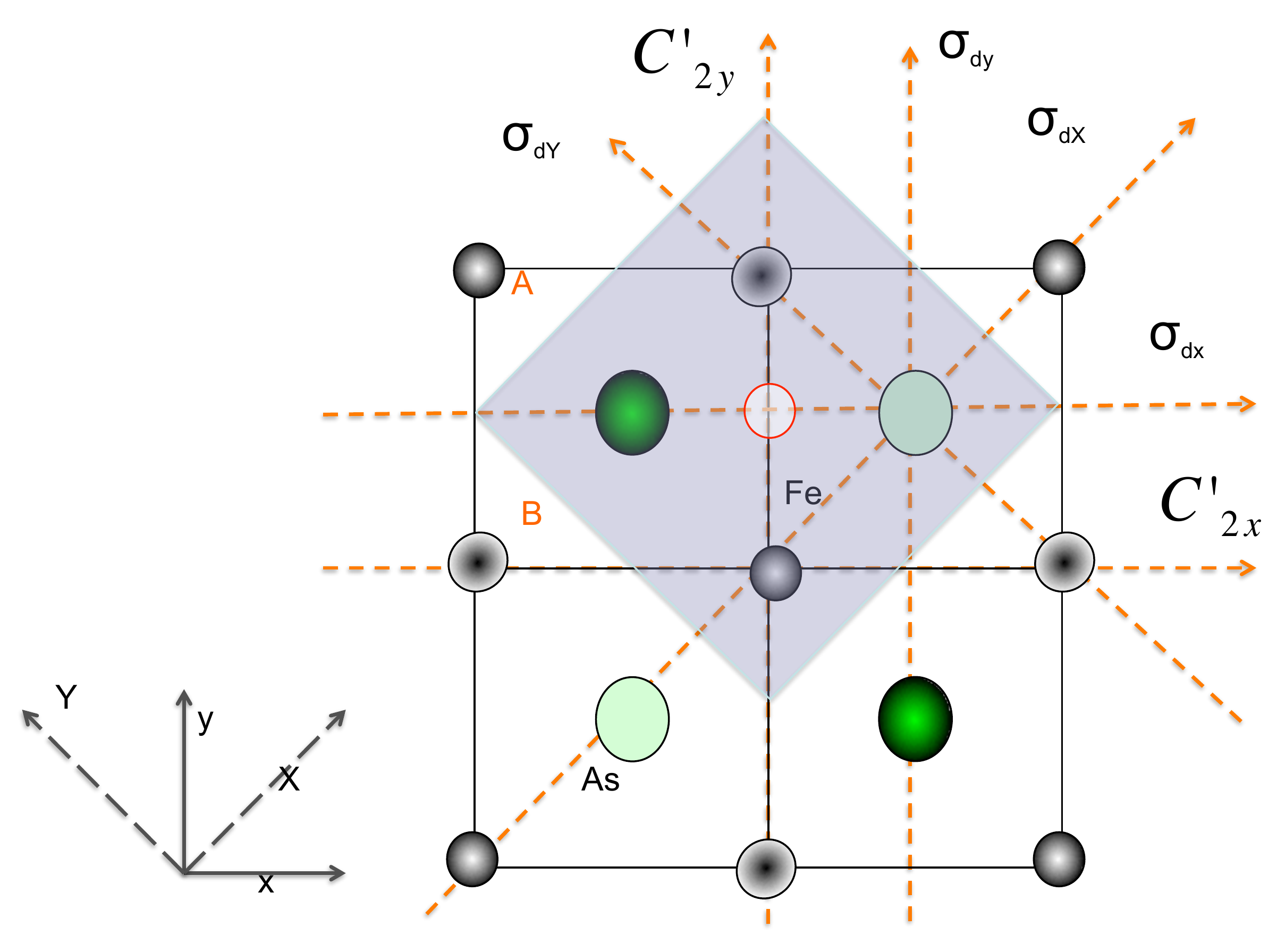}
\end{center}
\caption{Sketch of the lattice structure  of a trilayer Fe-As(Se) unit.  Notations used in the paper for axis directions, reflection symmetries and two sublattices are noted. }\label{lattice}
\end{figure}

\section{Parity and  pairing symmetry classification}  The pairing symmetry of a translation invariant superconducting state is classified by the IRs of $G/T$. If we ignore spin-orbital coupling, the ground state is expected to be a parity eigenstate.    Since spin-singlet pairing  is overwhelmingly supported experimentally in iron-based superconductors\cite{Johnstonreview,review-hirschfeld2011}, we focus on spin-singlet pairing states.

Conventionally,  for a spin-singlet pairing state,  the parity   is even. However, due to the unit cell doubling, the parity operation here essentially takes mapping between two  sublattices, A and B, in the iron square lattice  as shown in Fig.1.
Therefore there is no parity constraint for the pairing within each sublattice governed by  $D_{2d}$.  Thus,  for each irreducible representation of $D_{2d}$, there are two different pairing states with opposite parities. 
The IRs of $G/T$ are direct product of the IRs of two subgroups.

The character table of the $D_{2d}$ group is shown in Table.\ref{character_c4v}.  There are five different IRs with four being one-dimensional, A and B,   and one being two-dimensional, E. 
 \begin{table}
\begin{tabular}{|c|c|c|c|c|c|c|c|c|c|}
\hline 
& E & $2S_4$ & $C_2(z)$ & $2C'_2$ & $2\sigma_d$ & Linear, rotations & Quadratic \\ 
\hline 
$A_1$ & 1 & 1 & 1 & 1 & 1 && $x^2+y^2$,$z^2$ \\ 
\hline 
$A_2$ & 1 & 1 & 1 & -1 & -1& $R_z$ &   \\ 
\hline 
$B_1$ & 1 & -1 & 1 & 1 & -1 &  & $x^2-y^2$ \\ 
\hline 
$B_2$ & 1 & -1 & 1 & -1 & 1 & z & xy \\ 
\hline 
E & 2 & 0 & -2 & 0 & 0 & (x,y) $(R_x,R_y)$ & (xz,yz) \\ 
\hline 
\end{tabular} 
\caption{Character table for $D_{2d}$ point group. }\label{character_c4v}
\end{table}
Let's consider the symmetry operation $\hat S_{4}^2$ in $D_{2d}$.   It is easy to show that  
\begin{eqnarray}
\hat I \hat S_{4}^2 = (\hat \sigma_h, \hat t'_2),
\label{ea1}
\end{eqnarray}
where $\hat \sigma_h$ is the reflection along z-axis and $\hat t'_2$ is an in-plane translation  by $(1,0,0)$, namely one iron-iron lattice distance.  Eq.\ref{ea1} leads to an extremely important conclusion: The parity is determined by the eigenvalues of operation  ($\hat \sigma_h, \hat t'_2$). It  is equal to or opposite to the eigenvalues for  one-dimensional (A and B)  or   two-dimensional  (E) IRs respectively because $\hat S_4^2=1$ for one-dimensional IRs and $\hat S_{4}^2=-1$ for two-dimensional IRs.   More specifically,    this conclusion leads to that  the pairing is translation invariant with respect to 1-Fe unit cell in the A or B state when the parity is even  and in the E state when the parity is odd.  The $\eta$-pairing takes place in the A or B state with  odd parity  and in the E state with  even parity.  

The above classification is independent of the number of orbitals and orbital characters as long as we have the gauge setting specified for $d_{xz}$ and $d_{yz}$ mentioned earlier. 

 It is  also important to note that the classifications with respect to $D_{2d}$ at iron sites or $C_{4v}$ at the center of iron squares  are equivalent in a sense that they can be mapped to each other.   $C_{4v}$ has the same number and type of IRs as $D_{2d}$.  
We notice the following important relation,
\begin{eqnarray}
\hat S_4^3=\hat I\hat C_4.
\end{eqnarray}
where $\hat C_4$ is the $\pi/2$ rotation operation in $C_{4v}$. For one-dimensional IRs in $D_{2d}$, the above equation reduces to $\hat S_4=\hat I \hat C_4$.  Therefore, for parity even pairing, namely normal pairing,   there is no difference whether states are classified according to $D_{2d}$ or $C_{4v}$ since $\hat S_4= \hat C_4$.  Namely, a normal pairing state has the same IRs with respect to both $C_{4v}$ and $D_{2d}$.  For parity odd $\eta$-pairing,   $\hat S_4= -\hat C_4$, which implies that an s-wave state classified by $A$-IRs in $D_{2d}$ must become an d-wave state classified by $B$-IRs  in $C_{4v}$.  For example, an $\eta$-pairing $A_1$ s-wave state classified in $D_{2d}$  belongs to  the $B_2$ d-wave  in  $C_{4v}$. 
Therefore,  for an $\eta$-pairing parity odd state,  the name of the state depends on how it is classified. For above example, one can either name  the $\eta$-pairing state  as $B_{2u}$ d-wave or $A_{1u}$ s-wave, depending on the classification point groups $C_{4v}$ or $D_{2d}$ respectively.

An odd parity superconducting state must have a sign change between the top and bottom As/Se layers.  However, this information is hidden in an effective model with only d-orbitals constructed on an iron square lattice.  From above symmetry analysis, we can track the parity information simply using $\hat \sigma_h$. Although $\hat \sigma_h$ is not a symmetry operation for a single Fe-As(Se) trilayer,  the $\eta$-pairing state can be viewed  as a state with \emph{an internal  negative iso-spin defined by $\hat\sigma_h$}.  For any normal pairing, $\hat \Delta^n$ and $\eta$-pairing $\hat \Delta^\eta$ order parameters which belong to   one-dimensional IRs of $D_{2d}$,  we have
\begin{eqnarray}
& & \hat\sigma_h \hat\Delta^n \hat \sigma_h =\hat \Delta^n \\
& & \hat \sigma_h\hat \Delta^{\eta}\hat \sigma_h=-\hat \Delta^\eta
\label{normal-eta}
\end{eqnarray}

We can extend above discussion for order parameters in a bulk material.
There are two different lattice structures along c-axis in iron-based superconductors, 11-type ( which includes 111($NaFeAs$) and 1111($LaOFeAs$) structures) and 122-type where the 11-type is translation invariant along c-axis while the 122-type is not.  

For the  11-type,  we can simply extend above analysis to  the nearest-neighbour (NN) inter-layer pairing.  For an even parity order parameters, we can have two possible terms:  $(\vec k, -\vec k)$ pairing which is proportional to $cos(k_z)$ and the $\eta$-pairing  $(\vec k, -\vec k+Q)$  which is proportional to $i sin(k_z)$.   For an odd parity order parameters, the two possible terms become  $(\vec k, -\vec k)$ pairing which is proportional to $i sin(k_z)$ and the $\eta$-pairing  $(\vec k, -\vec k+Q)$ which is proportional to $cos(k_z)$. Both terms can be in a same irreducible representations of $D_{2d}$.   

For the 122-type, the situation is rather different because the 122 structure has a symmorphic space group $I4/mmm$ with a point group $D_{4h}$ centered in the middle of two NN Fe-As(Se) layers. The translation symmetry is specified by $(1,1,0)$, $(1,0,1)$, and $(0,1,1)$. The space inversion and $\hat \sigma_h$ have  identical characters in any one-dimensional IRs.  A state with odd parity which belongs to one-dimensional IRs must have node lines on Fermi surfaces when $k_z=0$. Therefore, we are only allowed to construct even parity states.  For the intra-layer pairing, there are two different even parity order parameters: one is constructed by  $(\vec k, -\vec k)$ pairing and the other is constructed by  $(\vec k, -\vec k+Q_3)$ pairing where $Q_3=(\pi,\pi,\pi)$, namely the $\eta$-pairing.  The difference between these two order parameters is that the latter  breaks $(1,0,1)$ and $(0,1,1)$ translation symmetry. Now if  we consider the NN inter-layer pairing,  we can have two terms which are parity even and keep the translation symmetry: the normal pairing  $(\vec k, -\vec k)$    which is proportional to $cos(k_z)$ and   $(\vec k, -\vec k+Q)$ pairing  which is proportional to $i sin(k_z)$. Here we refer $Q=(\pi,\pi,0)$.  For the $(\vec k, -\vec k+Q_3)$  $\eta$-pairing,  there are  also two terms in the NN inter-layer  pairing: the $\eta$-pairing  $(\vec k, -\vec k+Q_3)$   proportional to $cos(k_z)$ and   $(\vec k, -\vec k+Q)$ pairing  proportional to $cos(k_z)$. Therefore, if the inter-layer pairing is  included, the superconducting state generally breaks  translation symmetry of the iron-square lattice.

\section{ Effective Hamiltonian and Hidden Symmetry}
The above symmetry analysis is based on the original lattice symmetry. As we mentioned above,  an effective model based on d-orbitals appears to have a different symmetry.  In the past studies, we treated the model in 1-Fe unit cell with a   $D_{4h}$ point group at iron sites.  The treatment, in general, violated the fundamental spirit of symmetry principle and might have resulted in fundamental errors. To pay a full respect to symmetry principle, we must understand the symmetry properties of the effective model under the original lattice symmetry.

 We consider a general Hamiltonian in a single trilayer  $Fe-As(Se)$ structure coordinated by Fe and As(Se) atoms, 
\begin{eqnarray}
\hat H= \hat H_{dd} +\hat H_{dp} + \hat H_{pp}+\hat H_I
\end{eqnarray}
 where  $\hat H_{dd}$, $H_{dp}$ and $\hat H_{pp}$ describe the direct hopping between two d-orbitals,  the $d-p$ hybridization between Fe and As(Se)  and the direct hopping between two p-orbitals respectively.  $\hat H_I$ describes any  standard  interactions.   Here we do not need to specify the detailed parameters.  This Hamiltonian has a full symmetry defined by the  non-symmorphic space group.
\begin{figure}
\begin{center}
\includegraphics[width=1\linewidth]{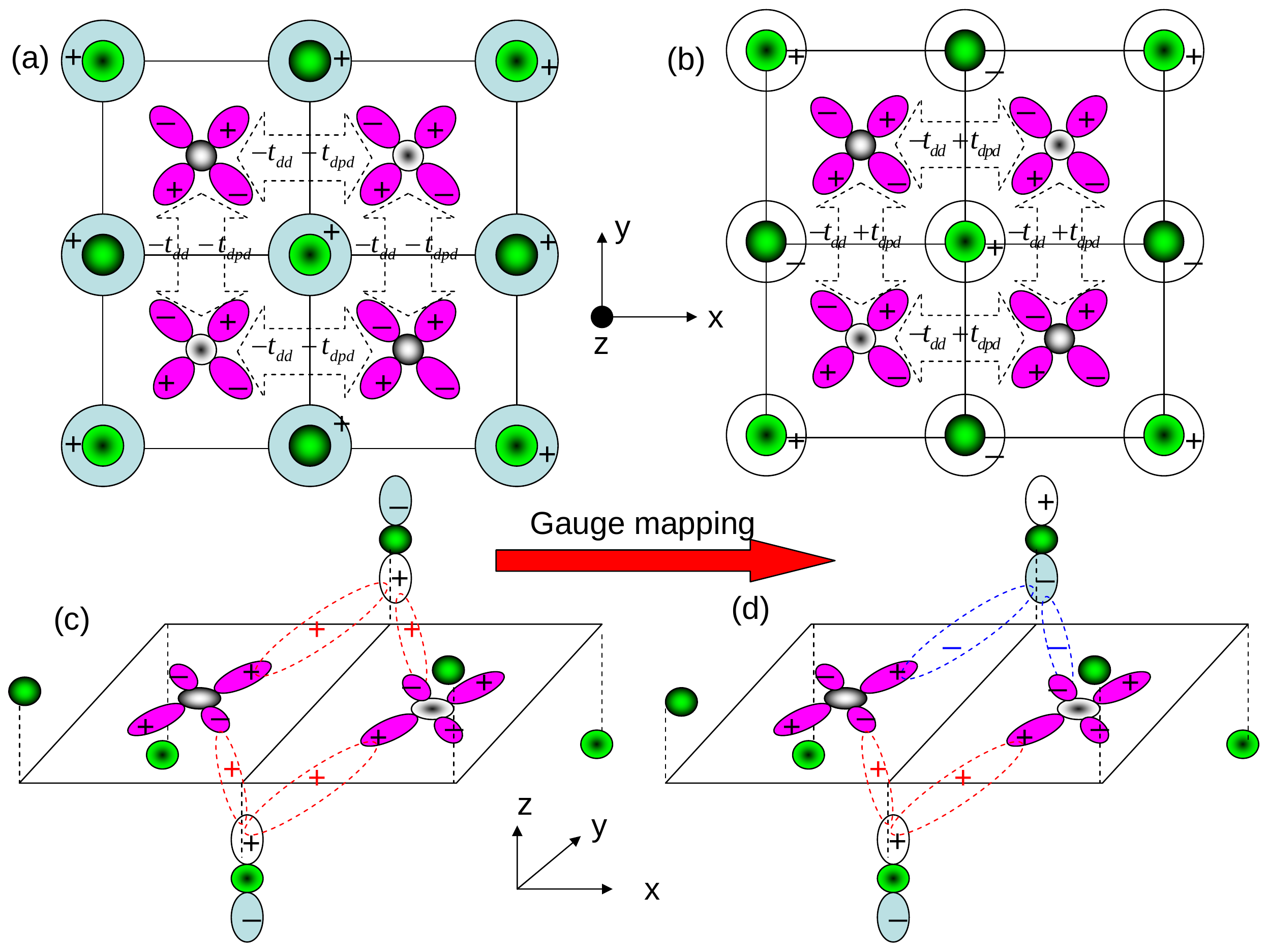}
\end{center}
\caption{(Color online) The nearest neighbor hopping parameters for intra-$%
d_{xy}$-orbital are shown in (a) and (b). $t_{dd}$ is the amplitude of the direct hopping of $%
d_{xy}$ orbital while $t_{dpd}$ is the amplitude of the indirect hopping through $p_{z}$
orbital of As/Se atom. The reason of sign change for $t_{dpd}$ between (a)
and (b) is that the $p_z$ orbitals in top layer and bottom layer  form
occupied bonding states in (a) and empty anti-bonding states in (b). The difference is illustrated by filled and empty $p_{z}$ orbitals in (a) and (b). (c) and (d) shows the local $p-d$ s-wave pairing
pattern  and the gauge transformation between them. }
\label{fig1}
\end{figure}

An effective Hamiltonian is obtained by integrating out  $p$-orbitals, which can be written as
\begin{eqnarray}
\hat H_{eff}= \hat H_{dd,eff} +\hat H_{I,eff} \label{eff}.
\end{eqnarray}
The effective band structure can be written as $\hat H_{dd,eff}= \hat H_{dd}+\hat H_{dpd}$, where $\hat H_{dpd}$ is  the effective hopping induced through d-p hybridization.  $\hat H_{dd,eff}$ has been obtained by many groups\cite{Kuroki2011,Miyake2009,Graser2009,Graser2010,Eschrig2009}. The major effective hopping terms  in $\hat H_{dpd}$ can be divided into two parts $\hat H_{dpd,NN} $,  which describes NN hopping and $\hat H_{dpd,NNN}$, which describes NNN hoppings in the iron square lattice. If one carefully checks the effective hopping parameters for  $t_{2g}$ orbitals in $\hat H_{dpd,NN}$,  one finds that they have opposite sign to what we normally expect  in a natural  gauge setting as shown in fig.\ref{fig1}(a,b), where $d_{xy}$ orbital is illustrated as an example. We see that  the hopping parameter $t_{dd}$ must be negative.  However   the effective hopping  parameter, $t_{dpd}$, is  positive and  even larger than $|t_{dd}|$ in \cite{Kuroki2011,Miyake2009,Graser2009,Graser2010,Eschrig2009}.  In a tetragonal lattice, $t_{dpd}$ can only be generated through  $d_{xy}-p_z$ hybridization. A positive value of $t_{dpd}$ suggests that  virtual hopping which generates $t_{dpd}$ must go through an unoccupied $p_z$ state. As shown in fig.\ref{fig1}(a,b),  a $d_{xy}$ equally couples to $p_z$  orbitals of top and bottom As atoms. A high energy $p_z$ state must be an anti-bonding $p_z$ state between NN As atoms. This analysis is held for all $t_{2g}$ orbitals which play the dominating role  in low energy physics.  It is also easy to check that the effective NNN hoppings between $t_{2g}$ orbitals are dominated through an occupied $p$ states, which is primarily a bonding state of $p$ -orbitals.  Therefore, the NN effective hoppings are generated through $d-p_a$ hybridization, where $p_a$ represents an anti-bonding p-orbital states and the NNN effective hoppings are generated through $d-p_b$ hybridization where $p_b$ is the bonding p-state.

\begin{figure}
\centerline{\includegraphics[width=1\linewidth]{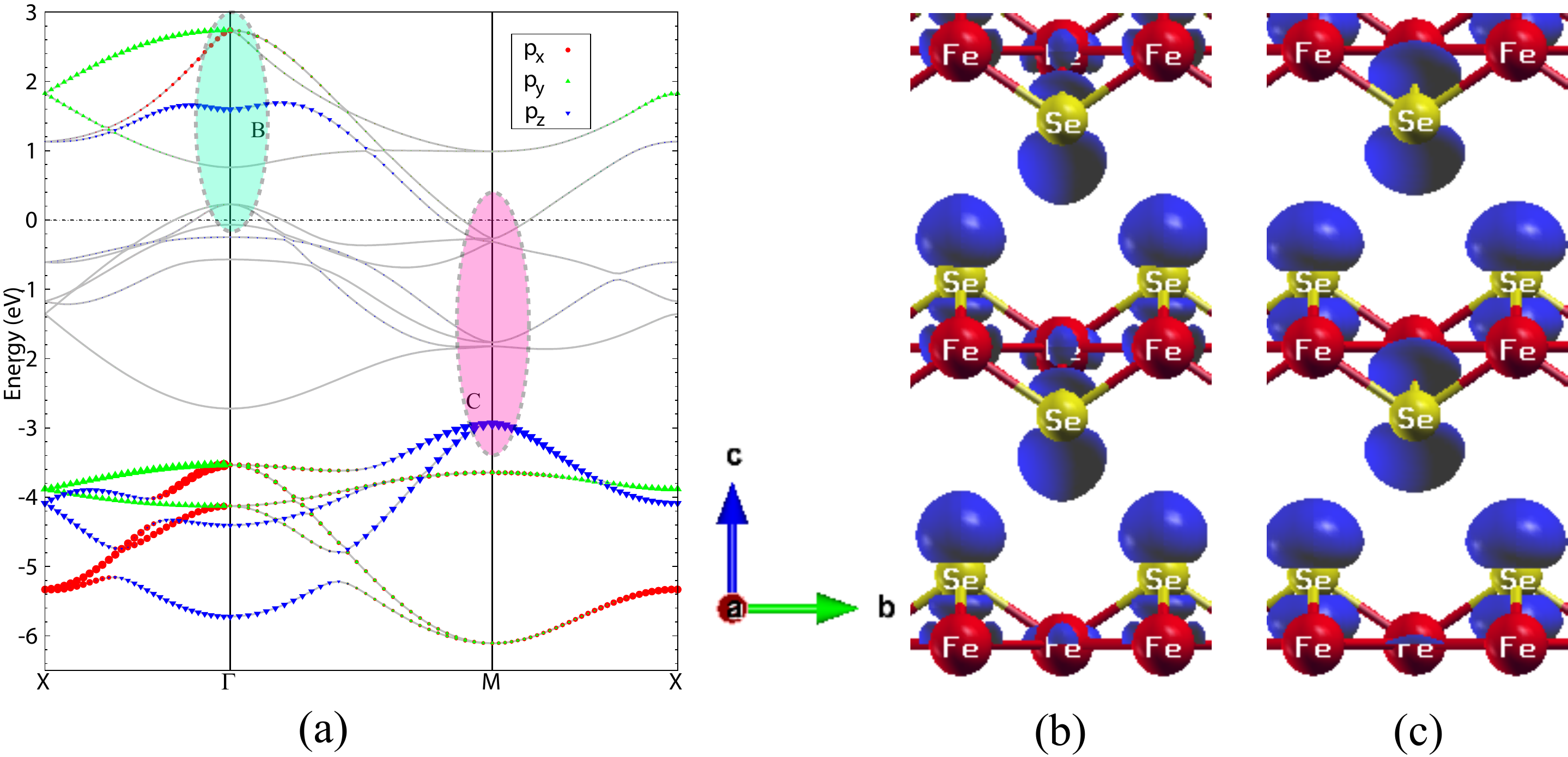}}
 \caption{ (a) The calculated band structure of  FeSe with the weight of p orbitals of Se. (b) The decomposed charge density  of  the band at $\Gamma$ marked by letter B (anti-bonding states). (c)The decomposed charge density  of  the band at $M$ marked by letter C (bonding states). 
 \label{fig2} }
\end{figure}

The above microscopic understanding is not surprising. In fact,  it is known in LDA calculations\cite{Miyake2009,Ma2008alu,singh2008a} that p-orbitals in As/Se are not fully occupied and there are significant overlappings between p-orbitals on bottom and top As/Se layers. Moreover, since $\hat H_{dpd,NN}$ and $\hat H_{dpd, NNN} $ primarily affect  hole pockets  around $\Gamma$ and electron pockets at $M$ separately, we can check the distribution of anti-bonding p states and bonding p states in  band structure to further confirm the analysis.   In fig.\ref{fig2}(a), we  plot the band structure of  FeSe and the distribution of $p$ orbitals. The $p_z$ orbitals of Se are mainly at +1.5 eV  at $\Gamma$ and -3 eV at $M$. By analyzing  the  bands at $\Gamma$ and $M$  as shown in Fig.\ref{fig2}(b) and (c), we confirm that  the   $p_z$ orbitals of Se at $\Gamma$ and $M$ belong to  anti-bonding and bonding states separately.

Knowing the above hidden microscopic  origins  in a derivation of an effective Hamiltonian allows us to understand the symmetry characters of the effective Hamiltonian in the original lattice symmetry. 
  
 The $d-p_a$ hybridization is odd under $ \hat \sigma_h$ while the $d-p_b$ hybridization is even under $ \hat \sigma_h$. Thus,  the NN hopping $\hat H_{dpd,NN}$ and NNN  hopping $\hat H_{dpd,NNN}$ should be classified as odd and even under $\hat \sigma_h$ respectively. Namely,
   \begin{eqnarray}
  & & \hat \sigma_h  \hat H_{dpd,NN}  \hat \sigma_h=-\hat H_{dpd,NN} \nonumber \\
  & &   \hat \sigma_h  \hat H_{dpd,NNN}  \hat \sigma_h=\hat H_{dpd,NNN} 
  \label{symmetry}
   \end{eqnarray}
The above hidden symmetry property   is against the main  assumption taken in many weak coupling approaches, which assume that the essential physics is driven by the interplay between hole pockets at $\Gamma$ and electron pockets  at $M$ \cite{review-hirschfeld2011}.  As indicated in fig.\ref{fig2}(a),  the interplay between the hole and electron pockets must be minimal because of their distinct microscopic origins. 

\section{ Gauge principle and Parity Conservation}
The symmetry difference in eq.\ref{symmetry} has fundamental impact on how to consider the parity  of a superconducting state if superconducting pairing is driven by local d-p hybridization.

  It has been shown that   in a system where  short range pairings in real space dominate, superconducting order parameters are momentum dependent and a gauge principle must be satisfied because the phases  of superconducting order parameters can be exchanged with those of the local hopping parameters\cite{bergkivelson2010,Hu2012s4} by gauge transformations.  As an example, a d-wave superconducting state in cuprates can be mapped to a s-wave superconducting state by a gauge mapping which changes the hopping terms from s-type symmetry to d-type symmetry\cite{Hu-s4review}. Therefore, only  the combined symmetry   of    hopping terms and  pairing orders associated to them  is a gauge-independent symmetry character to classify superconducting states. Namely, the symmetry of a superconducting state is characterized by
  \begin{eqnarray}
Symmetry_{sc}=  [\hat H_{hopping}][\hat \Delta]
  \end{eqnarray}
  where $[\hat A]$ indicates the symmetry of  $\hat A$.
  This gauge principle does not exist in a conventional BCS-type superconductor in which the information of pairing in real space is irrelevant. 
  
\begin{figure}
\begin{center}
\includegraphics[width=1\linewidth]{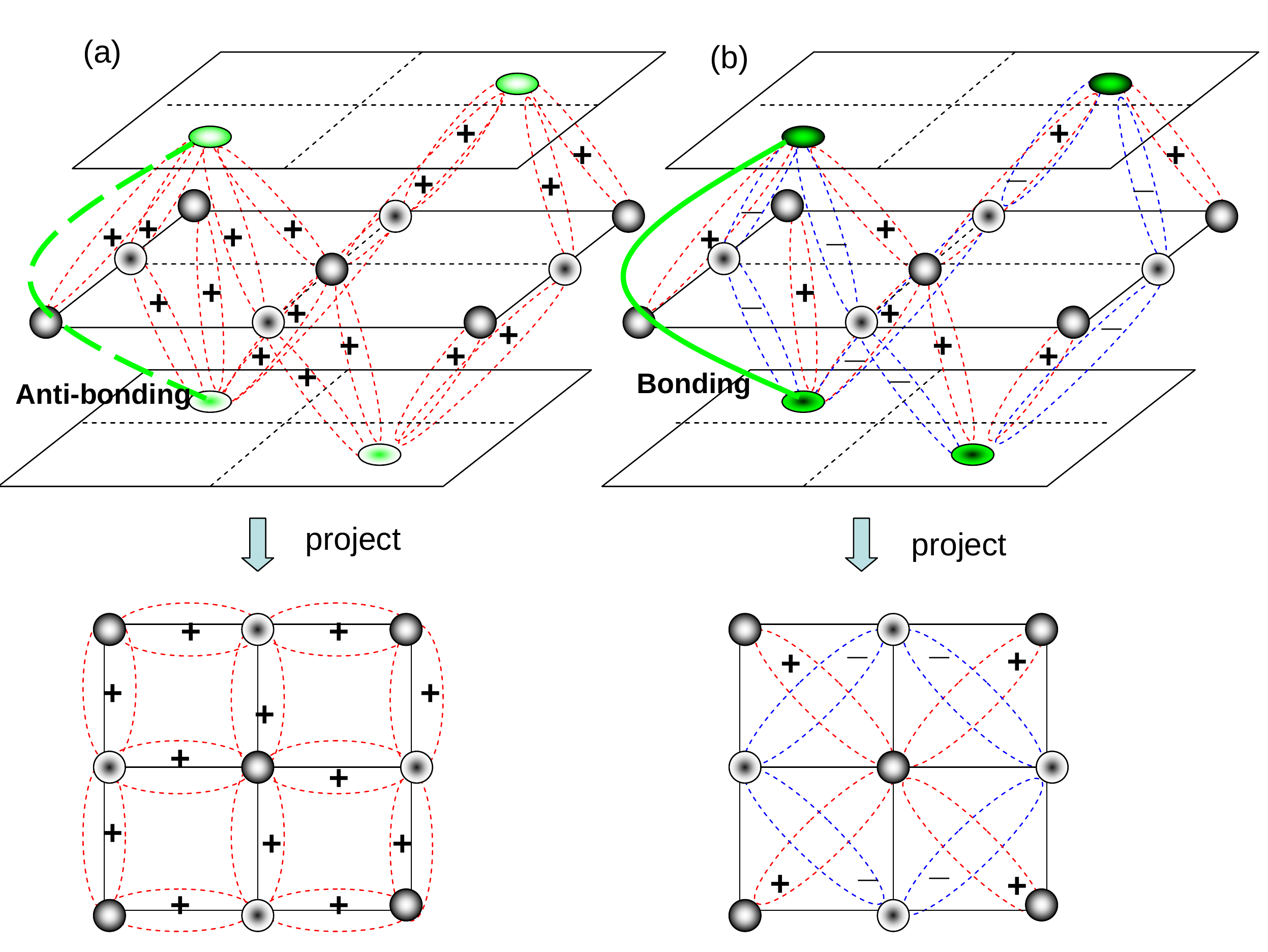}
\end{center}
\caption{(Color online) The NN and NNN $p-d$ local
pairing patterns with odd parity are shown in (a) and (b) in the natural gauge.  Note that p orbitals of As/Se in (a) form the anti-bonding
states while that in (b) form the bonding states. We distinguish the two
states with different filled green balls between (a) and (b). The $p-d$
pairings can be projected into effective $d-d$ pairings shown in the bottom
row. }
\label{fig3}
\end{figure}

Now we apply the gauge principle and  let $\hat \Delta_{NN}$ and $\hat \Delta_{NNN}$ be superconducting order operators associated with  $\hat H_{dpd,NN}$ and $\hat H_{dpd,NNN}$ respectively. In a superconducting state which belongs to a pure IR of the original lattice symmetry, we must have 
\begin{eqnarray}
[\hat \Delta_{NN}][\hat H_{dpd,NN}]=[\hat \Delta_{NNN}] [\hat H_{dpd,NNN}]
\label{gauge}
\end{eqnarray}

If we consider a superconducting state which conserves parity,
 following eq.\ref{symmetry}, we  have \begin{eqnarray}
  [\hat \Delta_{NN}]=-[\hat \Delta_{NNN}] \end{eqnarray}
   under $\hat \sigma_h$.  Therefore, based on the classification of pairing symmetries in Eq.\ref{normal-eta},   a parity conserved superconducting state must be a combination of normal pairing and $\eta$-pairing. 
   If $\hat \Delta_{NN}$ is a normal pairing, we immediately conclude that the state is parity  odd   and the superconducting order $<\hat \Delta_{NNN}>$ must be an $\eta$-pairing.

The above analysis can be easily illustrated in real space. As shown in fig.\ref{fig3}, if superconducting pairing is driven by local d-p hybridization, the superconducting order  is  a pairing between d and p orbitals $\Delta_{dp}= <\hat d^+\hat p^+>$. A uniform $ <\hat d^+\hat p_a^+>$ is parity odd.  The NN pairing, $<\hat \Delta_{NN}>$ in the effective model, must originate from $ <\hat d^+\hat p_a^+>$ and thus is also parity odd.  The gauge principle  can be understood as shown in fig.\ref{fig1}(c,d).  If we can take a new gauge for Fermion operators of p-orbitals, $\hat p\rightarrow -\hat p $, in one of  the two As(Se) layers,   the anti-bonding operator $\hat p_a$ maps to the bonding operator $\hat p_b$. This gauge mapping  exactly transfers  the parity  between hopping terms and superconducting order parameters.

 \section{ Meanfield Hamiltonian for parity conserved s-wave state}
 The above analysis can be generalized to all effective hoppings. The basic idea is to divide the iron square lattice into two sublattices.  
 In an odd parity  state, the pairing between two sublattices must be normal pairing while the pairing within sublattices must be $\eta$-pairing.  In an even  parity state, the pairing between two sublattices must be $\eta$-pairing while the pairing within sublattices must be normal pairing.  This means that the pairing between two sublattices must be vanished in an even parity state.
 
 In all of measured samples of iron-based superconductors, no  universal node along $\Gamma-M$  and $\Gamma-X$ directions on all Fermi surfaces were  observed\cite{arpes-zhao2008,arpes-ding2008a,arpes-wray2008,arpes-ding2008,Lisy2009,arpes-nakayama2010,Feng2011,arpes-zhang2011feng,Wangxp2012,Xum2012}.  These experimental facts place a constraint  that the superconducting state must be in the $A_1$ IR of  $D_{2d}$, namely, an s-wave viewed at iron sites.  Then, the remaining question is about the parity of the state.
 
  An even parity  s-wave state, the normal pairing between two sublattices must be vanished. Therefore, the meanfield Hamiltonian for an even parity s-wave state is
   \begin{eqnarray}
 H^{e}_{mf}= H_{dd,eff}+\sum_{\alpha,\beta, k}( \delta^e_{\alpha\beta, n} \hat \Delta_{\alpha\beta,n}(\vec k)+h.c.)
\label{mf}
 \end{eqnarray}
where  $\hat \Delta_{\alpha\beta,n}= \hat d_{\alpha\uparrow}(\vec k)\hat d_{\beta\downarrow}(-\vec k)-d_{\alpha\downarrow}(\vec k)\hat d_{\beta\uparrow}(-\vec k) $ and $\alpha,\beta$ label orbital.  In general, the normal pairing order parameters must satisfy
\begin{eqnarray}  
\delta^e_{\alpha\beta, n} (\vec k)= \delta^e_{\alpha\beta,n}(\vec k+Q) 
\label{even-mf}
\end{eqnarray} 
All superconducting states derived before from weak coupling approaches were considered as even parity\cite{Mazin2008, Kuroki2011,wangfa2009,Graser2010,Chubukov2010,thomale,review-hirschfeld2011}. However,   as a normal pairing between two sublattices is included, the parity is not conserved. In strong coupling models\cite{Seo2008,Lu2012,sc-fang2011,yurong2011},  the superconducting order was derived from NNN antiferromagnetic  exchange coupling $J_2$, which is a normal pairing within sublattices, satisfies 
Eq.\ref{even-mf}.    Therefore, if  superconductivity is only originated from $J_2$, the proposed state is an even parity s-wave state, namely, the $A_{1g}$ s-wave.

This state provides a good understanding of superconducting gaps in iron-based superconductors. However, it can not explain dual symmetry characters with both s-wave and d-wave type observed in the superconducting state, for example,  the spin relaxation $\frac{1}{T_1T}$ measured by nuclear magnetic resonance(NMR).  A coherent peak around $T_c$  is expected in a full gap s-wave state even if it is a $s^{\pm}$\cite{review-hirschfeld2011,Parish2009}.  Experimentally,  in  an extremely clean sample where the exponential temperature dependence was measured,   no coherent peak was observed yet at $T_c$\cite{nmr-li2011}.  For iron-chalcogenides\cite{ Guojg2010, Hesl2012,Liudefa2012,Tansy2013}, this state  is highly questionable because of the absence of sign change on Fermi surfaces so that it is hard to explain the possible sign change evidence from neutron scattering\cite{resonance-park2011}.  It is also worth mentioning that  
many weak coupling methods suggest that the superconducting state in iron-chalcogenides is a d-wave with normal pairing\cite{Maier2011,Khodas2012,Wangfa2011}.  This state  is not consistent with  experimental results showing  the absence of  nodes on high symmetry lines\cite{arpes-zhang2011feng,Wangxp2012,Xum2012} and the presence of strong ferromagnetic NN exchange coupling \cite{spinwave-wang2011a,Hu2012u}.  Nevertheless, this d-wave state should be considered as an odd parity state in the original lattice symmetry because it only includes normal pairing between two sublattices.

A meanfield Hamiltonian to describe the odd parity s-wave state in 1-Fe unit cell ( $A_{1u}$ s-wave) can be generally written as
 \begin{eqnarray}
 & & H^o_{mf}= H_{dd,eff}+\sum_{\alpha,\beta, k}( \delta^o_{\alpha\beta, n} \hat \Delta_{\alpha\beta,n}(\vec k)\nonumber \\
 & & +\delta^o_{\alpha\beta, \eta} \hat \Delta_{\alpha\beta,\eta}(\vec k)+h.c.)
\label{mf}
 \end{eqnarray}
where   $ \hat \Delta_{\alpha\beta,\eta}= \hat d_{\alpha\uparrow}(\vec k)\hat d_{\beta\downarrow}(-\vec k+Q)-\hat d_{\alpha\downarrow}(\vec k)\hat d_{\beta\uparrow}(-\vec k+Q)$.  In general, the normal   and $\eta$ pairing order parameters satisfy
\begin{eqnarray} & & \delta^o_{\alpha\beta, n} (\vec k)= -\delta^o_{\alpha\beta,n}(\vec k+Q) \\
& & \delta^o_{\alpha\beta, \eta} (\vec k)= \delta^o_{\alpha\beta,\eta}(\vec k+Q) .
\end{eqnarray}
These equations capture the sign change of superconducting order parameters in momentum space. The sign change here is required by odd parity symmetry.  

While detailed studies will be carried out in the future, the meanfield Hamiltonian captures superconducting gaps in both iron-pnictides and iron-chalcogenides. As the inter-orbital pairing can be ignored for s-wave pairing and the pairing is dominated by NN and NNN pairings, the important parameters are 
\begin{eqnarray}
& & \delta^o_{\alpha\alpha, n} \propto cosk_x+cosk_y\nonumber \\
 & & \delta^o_{\alpha\alpha, \eta} \propto cosk_xcosk_y.
 \end{eqnarray}
 Thus, the superconducting gaps on hole pockets are mainly determined by  $\delta^o_{\alpha\alpha, n} $  and those on electron pockets are mainly determined by $\delta^o_{\alpha\alpha, \eta}$. 
 
\begin{figure}
\begin{center}
\includegraphics[width=1\linewidth]{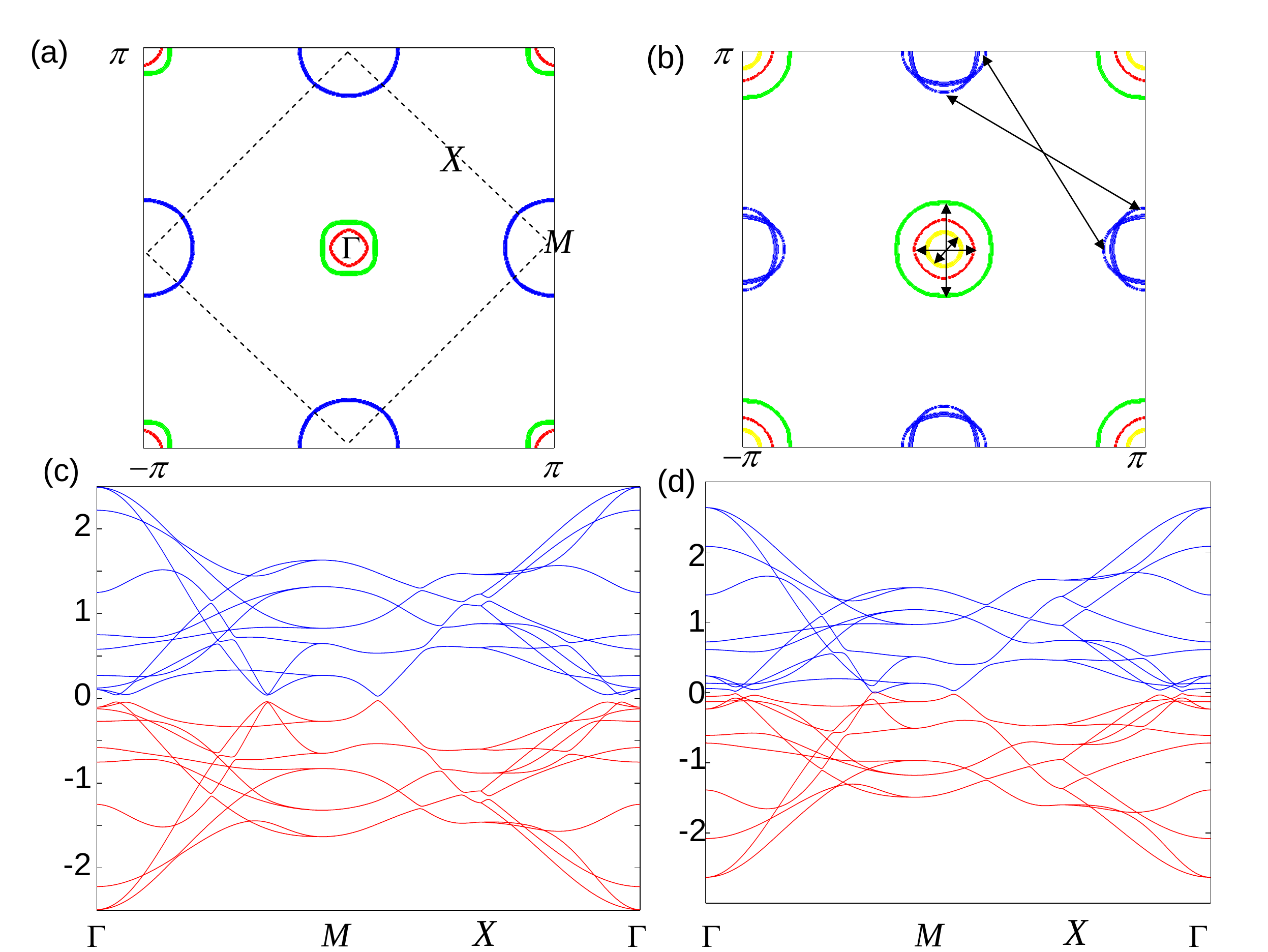}
\end{center}
\caption{(Color online) The Fermi surfaces of a five-orbital model in\cite{Graser2009}  are
shown in (a) and (b). The forms of hopping terms and hopping parameters can
be found  in\cite{Graser2009}. Here, we only add a chemical
potentials to tune the fermi level. We set $\protect\mu =0.1$ and $-0.04$ in
(a) and (b). The high-symmetry points are shown in (a) and the pairing
channels connecting the points on the fermi surface are denoted by the black
lines with arrows. The quasi-particle spectrum of the superconductive
states for (a) and (b) are shown in (c) and (d). We can find the (c) is full
gaped and (d) has nodes at the electron pockets. The superconductive order
parameters are chosen: $\Delta _{11,x}^{N}=\Delta _{11,y}^{N}=0.05;$ $\Delta
_{44}^{N}=0.05;$ $\Delta _{11}^{NN}=0.05;$ $\Delta _{12}^{NN}=0.05;\Delta
_{44}^{N}=-0.1;$ }
\label{fig4}
\end{figure}

  In the odd parity s-wave  state, there is no symmetry protected node. However, accidental nodes can easily take place.  In fig.\ref{fig4}, we plot numerical results for two cases.  Parameters are specified in the caption of the figure.   The superconducting gap  in the first case is a  full gap while it has gapless nodes on electron pockets in the second case.
This may provide an explanation why  gapless excitations were observed in some materials\cite{review-hirschfeld2011}. The detailed studies will be reported in future.

The odd parity s-wave state also explains  the dual symmetry characters of both s-wave and d-wave type in iron-based superconductors.   The $\eta$-pairing s-wave order in $D_{2d}$  essentially is a d-wave type order  according to $C_{4v}$ as shown in Fig.\ref{fig3}. For a d-wave pairing symmetry, the vanishing of the coherence factor is required by symmetry.  Thus, current NMR results really support the odd parity state.  

The Hamiltonian in Eq.\ref{mf} can not be reduced to a translationaly invariant Hamiltonian in 1-Fe unit cell. The  2-Fe unit cell  becomes  intrinsic.   Unique features related to 2-Fe unit cell  should  be observed and studied\cite{wku}. A detailed study will be carried out in future.

In summary, we provide the meanfield Hamiltonian for parity conserved superconducting states. 
Parity conservation was seriously violated in the past studies. Comparing odd and even parity s-wave states,  we show that the odd parity s-wave state can naturally explain many intriguing properties of iron-based superconductors.

\section{Signature of odd parity superconducting state}
The fingerprint  of the odd parity state is the negative iso-spin of $\hat \sigma_h$, which indicates the sign change of order parameters between top and bottom As(Se) layers. This property was first revealed in  the recent  constructed effective $S_4$ symmetry model based on the 2-Fe unit cell\cite{Hu2012s4,Hu-s4review,Ma2012-s4}.  However, as the $S_4$ model is   a simplified effective model based on two effective orbitals,  the parity characters were not revealed. 

\begin{figure}
\begin{center}
\includegraphics[width=1\linewidth]{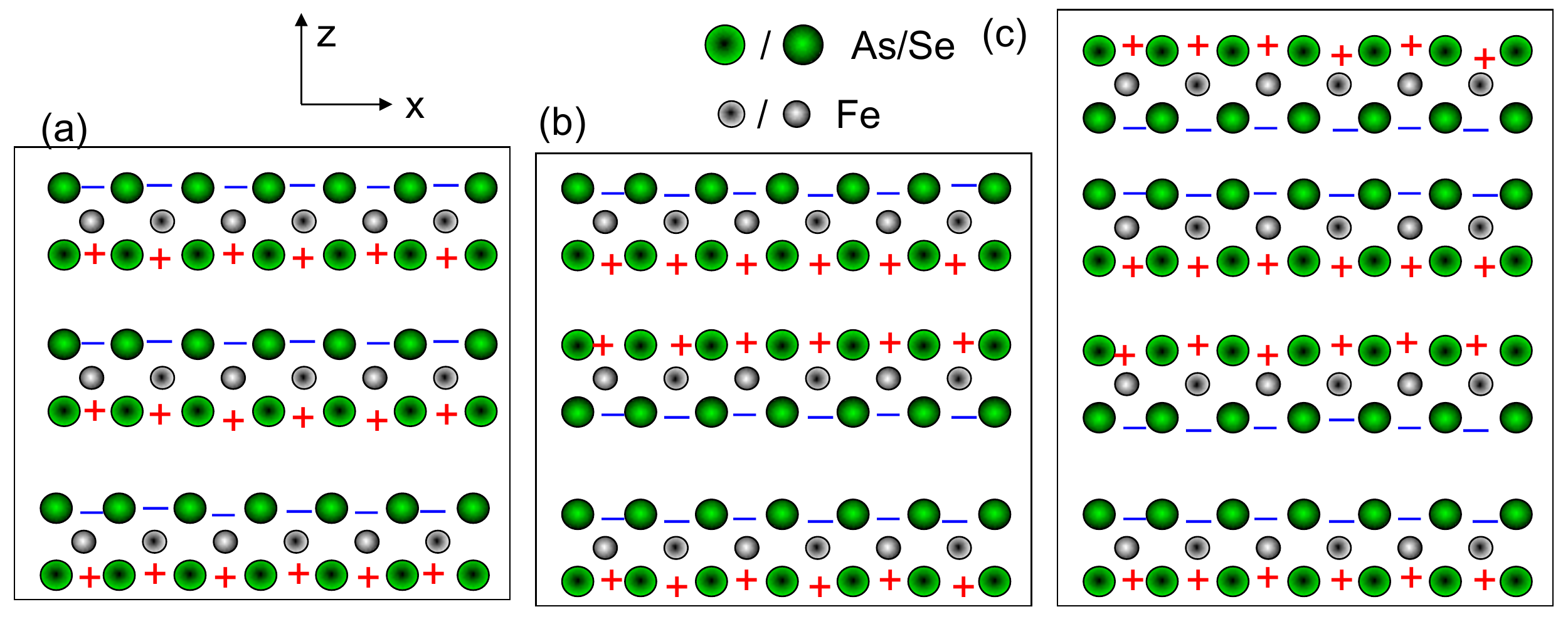}
\end{center}
\caption{Sketch of real space sign distribution  along c-axis  for an $\eta$-pairing  state in  (a) 11-type structure,  (b)  122-type structure with odd number of layers  and (c) 122-type structure with even number of layers. }\label{sign}
\end{figure}

 The odd parity  indicates  that a superconducting single Fe-As(As) trilayer is a $\pi$-junction along c-axis, which can be measured   in a single crystal material as shown in Fig.\ref{sign}:  a uniform superconducting state is characterized by sign change  between top and bottom surfaces along c-axis in 11-type structure while in the 122-type structure, the sign change only presents when the number of layers are odd. 

\section{Discussion}
In the history of condensed matter physics,  a new quantum state of condensed matter  is hardly obtained by solving a model. Here we use fundamental principles to show that an odd parity  state can be naturally taken place in Iron-based superconductors and suggest smoking-gun experiments to detect or falsify it.  With microscopic understanding proposed here, a detection of the odd parity state  can have a tremendous impact on high $T_c$ mechanism  for  iron-based superconductors and other high $T_c$ superconductors.

In the past five years, many researches based on effective models suggest that pairing symmetries  in  iron-based superconductors  are very fragile.  Those studies essentially suggest that principles to understand  the robustness of superconductivity and pairing symmetry are still missed in our standard approach. The results in this paper  demonstrate that previous studies did not correctly  take  parity conservation and hidden symmetry in effective models into account and mishandled    symmetry  and gauge principles.  In the past,  the effective Hamiltonian was viewed  in the symmetry group $D_{4h}$ at iron sites rather than the original lattice symmetry $G$.  We can see that if $\hat \sigma_h$ could  be set to one, $G$ is equivalent to $D_{4h}$. However,  due to the  anti-bonding $p$ orbital states,  the effective Hamiltonian does not represent correct symmetry of the original lattice in a natural gauge setting.   It will be interesting to see how the missing pieces can be properly implemented in our standard methods.

Our results suggest that gauge principle needs to be properly implemented when we derive effective Hamiltonian to simplify a complex system. The correct physics can only  be understood after the hidden gauge is revealed. For  an order parameter which is momentum dependent,  this gauge information is critical.    The gauge principle becomes very important  for us to search new physics in other complex electron  systems.

The odd parity state also suggests the importance of correlated electron physics. It is believed that sign change superconducting order is inevitable in a superconducting state of strongly correlated electron systems because of the existence of strong repulsive interaction.  This principle is violated in a parity even s-wave state.  A measurement of the  party odd s-wave state  can provide a strong support of this principle.

The microscopic mechanism revealed here   fundamentally  differs from those proposed in weak coupling approaches which only emphasize Fermi surfaces. Fermi surfaces   are only determined by energy dispersion. It provides no information about   underlining microscopic processes which are  local and  bound with high energy physics.  In correlated electron systems, these processes essentially  determine many important properties. This is also the reason why iron-pnictides and iron-chalcogenides can be unified even if their Fermi surfaces are drastically different.

If the odd parity state is confirmed,  the fundamental objects  in   superconducting  states of high $T_c$ materials must be the  tightly binding Cooper pairs between d and p orbitals.  In this view, the odd parity s-wave state closely resembles the d-wave state in a Cu-O plane of cuprates. We expect an  identical mechanism to select sign changed superconducting orders in both materials.   From Fig.\ref{fig3}(b), one can see that the $\eta$-pairing part in the odd parity state can be viewed as two d-wave states formed in two sublattices, a direct analogy to the d-wave in cuprates.

This study opens a promising new direction for the research in iron-based superconductors and suggests that the physics  in these materials  is deeper  and much more inspiring than what we realized before.  It  leaves many unanswered questions for us. A few questions are in order. First, what is relationship between magnetism and superconductivity?  One can see that the collinear antiferromagnetic  (C-AFM)  state\cite{daihureview} observed in iron-pnictides has odd parity. This may partially answer why superconductivity and C-AFM order can coexist in the phase diagram.  
Second, what are other unique properties in an odd parity s-wave state? It is known that an odd parity p-wave state displays many unique properties.  Third, what is the relationship between nematism and superconductivity? Nematism  breaks rotational symmetry and was observed at high temperature\cite{nematic-fisher2011}. Fourth, how robust is an odd parity state in response to impurity?   Finally, this study points out that there are three possible scenarios related to parity in superconducting states, even, odd or broken. Experiments proposed here  will finally nail down  the truth.  All the previous studies took even parity for granted without knowing  that the parity was acturally broken in the proposed states. However, a parity breaking superconducting state is also  an interesting state to explore.

In summary,  using symmetry and gauge principles, we show that  iron-based superconductors are unified into an odd parity s-wave superconducting state.  We demonstrate that  in an effective model based on d-orbitals,  superconducting states studied in the past violate parity conservation.  The existence of the odd parity state can have tremendous impact on  high $T_c$ superconducting mechanism.

{\it Acknowledges:}  The author acknowledges   NN Hao for useful discussion and  providing the numerical results for superconducting gaps  and XX Wu for providing the figure of the distribution of p-orbital in LDA calculations.  The author wants to thank  for   H. Ding, P. Coleman, D. Scalapino, T. Xiang, X. Dai,  and D.L. Feng for useful discussion.  The work is supported  by the Ministry of Science and Technology of China 973 program(2012CB821400) and NSFC-1190024.

\end{document}